\documentclass[aps,prc,twocolumn,showpacs,amsmath,amssymb,preprintnumbers]{revtex4-1}
\usepackage[bookmarks=false]{hyperref}
\usepackage{latexsym} % Gets \Box etc
\usepackage{tikz}
\usepackage{subfigure}
\usepackage{enumerate}
\usepackage{graphicx}       % For PostScript figures
\usepackage{bbm}       % For unit matrix symbol
\usepackage{color}
\usepackage{bm}   % Bold math: \bm{1}
\newcommand{\al}{\alpha}

\newcommand{\ga}{\gamma}

\newcommand{\de}{\delta}
\newcommand{\De}{\Delta}

\newcommand{\eps}{\epsilon}

\newcommand{\la}{\lambda}

\renewcommand{\th}{\theta}   % LaTeX: \th already defined

\newcommand{\Om}{\Omega}

\newcommand{\beq}{\begin{equation}}
\newcommand{\eeq}{\end{equation}}
\newcommand{\ba}{\begin{array}}
\newcommand{\ea}{\end{array}}
\newcommand{\bea}{\begin{eqnarray}}
\newcommand{\eea}{\end{eqnarray}}
\newcommand{\bi}{\begin{itemize}}  %\setlength{\itemsep}{0\parsep}}
\newcommand{\ei}{\end{itemize}}
\newcommand{\ben}{\begin{enumerate}} %\setlength{\itemsep}{0\parsep}}
\newcommand{\een}{\end{enumerate}}
\newcommand{\bc}{\begin{center}}
\newcommand{\ec}{\end{center}}

\newcommand{\<}{\langle}
\renewcommand{\>}{\rangle} % LaTeX: \> already defined
\newcommand{\txt}{\textstyle}

\newcommand\eqn[1]{(\ref{#1})}      % parentheses around the LaTex "ref" macro
\newcommand{\third}{{\txt \frac{1}{3}}}

\newenvironment{tightlist}[1]{ 
\begin{list}{#1}{
  \usecounter{enumi}
  \setlength{\topsep}{0ex} % added to \parskip
  \setlength{\itemsep}{-\parsep} % added to \parsep
  \settowidth{\labelwidth}{#1} % width of item label
  \setlength{\labelsep}{0.5em}        % space between label and text
  \setlength{\leftmargin}{\labelwidth}% margin = width + separation
  \addtolength{\leftmargin}{\labelsep}
 }}{\end{list}}
\newcommand{\Tr}{{\rm Tr}}
\newcommand{\Lagr}{\mathcal{L}}
\newcommand{\Hamil}{\mathcal{H}}

\usepackage[normalem]{ulem}  % \sout{old text} for strikeout

\begin{document}

\title{Stability of superfluid vortices in dense quark matter}
\author{Mark G.~Alford, S.~Kumar Mallavarapu} 
\affiliation{Physics Department, Washington University,St.~Louis, MO~63130, USA}
\author{Tanmay Vachaspati}
\affiliation{Physics Department, Arizona State University, Tempe, AZ~85287, USA}
\author{Andreas Windisch}
\affiliation{Physics Department, Washington University,St.~Louis, MO~63130, USA}
\pacs{12.38.-t, 25.75.Nq}

\begin{abstract}
Superfluid vortices in the color-flavor-locked (CFL) phase of dense quark
matter are known to be energetically disfavored relative to
well-separated triplets of ``semi-superfluid'' color flux tubes. However, the
short-range interaction (metastable versus unstable) has not been established.
In this paper we perform numerical calculations using the effective theory of
the condensate field, mapping the regions in the parameter space of coupling
constants where the vortices are metastable versus unstable. For the case of
zero gauge coupling we analytically identify a
candidate for the unstable mode, and show that it agrees well with the
results of the numerical calculations.  We find that in the region of the
parameter space that seems likely to correspond to real-world CFL quark matter
the vortices are unstable, indicating that if such matter exists in neutron
star cores it is very likely to contain semi-superfluid color flux tubes
rather than superfluid vortices.
\end{abstract}

\date{16 Jan 2016} % Hardwire date so arXiv does not change it

\maketitle
%\tableofcontents
%============================================================
%  I: INTRODUCTION 
%============================================================
\section{Introduction}
\label{sec:intro}
The densest phase of matter according to standard model physics
is the color-flavor-locked (CFL) phase \citep{Alford:1997zt}.
The CFL condensate breaks the baryon number symmetry of the theory,
hence the CFL phase is a superfluid, and CFL matter in the core of a spinning neutron star will carry angular momentum in the form of superfluid vortices.  
Unlike the fermions in terrestrial superfluids, quarks interact via a
non-Abelian gauge group; the structure of the vacuum manifold \citep{Balachandran:2005ev} is 
$SU(3) \times U(1) / Z_{3}$ and this permits the
existence of a non-Abelian vortex configuration which is three times lower in energy than the usual superfluid vortex. This configuration consists of three widely-separated semi-superfluid flux tubes, each carrying color magnetic flux.  At separations much larger than the size of the core any two semi-superfluid flux tubes strongly repel each other 
\citep{Nakano:2007dq}, and it has therefore been conjectured
that CFL superfluid vortices will spontaneously decay into triplets of
semi-superfluid flux tubes. However, the short-range interaction between
the flux tubes has not been calculated \citep{Eto:2013hoa}, 
leaving open the possibility
that the vortices might be metastable, in which case the decay rate,
occurring via barrier penetration, could be extremely low.
Such metastability has already been established for vortices in an
analogous system, a three component Bose-Einstein condensate \citep{Cipriani:2013wia}.

In this paper we address these unresolved questions. To probe the stability
of the CFL superfluid vortices we solve the classical field equations for the
CFL condensate on a two-dimensional lattice, analogously to previous
calculations done for SU(2) Yang-Mills-Higgs theory \citep{Gleiser:2008dt},\citep{Heinz:1996wx}. This approach gives us a full understanding of the decay
process, far exceeding the insight that can be gained from asymptotic methods.
The results presented in this paper are:
\begin{tightlist}{$\bullet$}
\item We map the regions in the parameter space of the couplings where the superfluid vortex is metastable (Section~\ref{sec:results}).
\item In the unstable regions we numerically extract the unstable mode (Section~\ref{sec:results}).
\item We analytically construct an unstable mode arising in the case of zero gauge coupling (Section \ref{sec:stability}), and show that it is very similar to the numerically extracted unstable mode.
\item We clarify the nature of interaction of semi-superfluid vortices at short distances in the unstable region.  
\end{tightlist}

\iffalse
In Section \ref{sec:vortex_solutions} we discuss the vortex solutions of the Ginzburg-Landau model. The stability of the superfluid vortex solution is then addressed in Section \ref{sec:stability}, where we construct an analytic Ansatz for the unstable mode in the case of zero gauge coupling. In Section \ref{sec:results}, we confirm the Ansatz numerically by solving the equations of motion on a lattice. Using this technique, we explore the parameter space of couplings to determine regions of (meta-) stability beyond the zero gauge coupling case. Finally, we summarize our findings in Section \ref{sec:conclusion}.
\fi

%============================================================
%  II: VORTEX SOLUTIONS - GENERAL
%============================================================
\section{Vortices and  flux tubes}
\label{sec:vortex_solutions}
The non-Abelian Ginzburg-Landau Lagrangian used to describe semi-superfluid vortices \cite{Balachandran:2005ev} in the CFL phase of dense quark matter is given by  
\bea
\Lagr &=& \Tr \left[-\frac{1}{4} F_{ij} F^{ij} + D_{i} \Phi ^{\dagger} D^{i} \Phi   + m^2 \Phi ^{\dagger} \Phi  - \lambda_{2} (\Phi ^{\dagger} \Phi)^2 \right] \nonumber \\ 
&& - \lambda_{1}(\Tr[\Phi^{\dagger} \Phi])^2 + \dfrac{3m^{4}}{4 (3\lambda_1+\lambda_2)} \ ,
\eea
where $D_{i} = \partial_{i} - i g A_{i}$, $F_{ij} = \partial_{i} A_{j} - \partial_{j} A_{i} - ig  \left[ A_{i},A_{j} \right] $. $A_{i}$ represents the gluonic gauge field. We choose the normalization $\Tr [T^{\alpha}T^{\beta}] = \delta^{\alpha \beta}$ for the $SU(3)$ generators. In the CFL phase $\Phi$ represents the color-flavor-locked diquark condensate. It is a  3$\times$3 complex matrix.
An element of $\Phi$ may be denoted as $\phi_{\alpha a}$, where $\alpha$ is a color index and $a$ is a flavor index. In the symmetry-breaking phase the field 
in the ground state has a non-zero vacuum expectation value,
\beq \label{eq:vev}
A_i = 0  \ , \quad  \Phi = \bar \phi \textbf{1}_{3 \times 3} \ ,
\quad \bar\phi = \sqrt{\frac{m^2}{2 \lambda}} \ ,
\eeq
where
\beq \label{eq:lambda_def}
\lambda \equiv 3 \lambda_{1} + \lambda_{2},  
\eeq
and the mass spectrum contains the Goldstone boson and two massive Higgs fields associated with perturbations along the singlet and adjoint directions. Their masses are $m_{\phi} = \sqrt{2}m$ and $m_{\chi} = 2 \sqrt{\lambda_{2}} \bar\phi$ respectively. Stability of the ground state requires $\lambda > 0 $ and $\lambda_{2} > 0$. 
There are also massive gluons of mass $m_{g} = \sqrt{2} g \bar\phi$.

The superfluid vortex is
\beq \label{eq:superfluid_solution}
A_i = 0  \ ,\quad
 \Phi_{\mathrm{sf}} = \bar\phi\, \beta(r) e^{i \theta} \,\textbf{1}_{3 \times 3} \ , 
\eeq
where $\beta(r)$ is the radial profile that satisfies
\beq \label{eq:radial_profile}
\beta^{''} + \frac{\beta^{'}}{r} - \frac{\beta}{r^{2}} - m^{2} \beta (\beta^{2} - 1) = 0 \ ,
\eeq
with boundary conditions
\beq
\ba{rcl}
\beta &\rightarrow& 0 \ \mathrm{as} \ r \rightarrow 0 \ , \\
\beta &\rightarrow& 1 \ \mathrm{as} \ r \rightarrow \infty \ .
\ea
\eeq 
However, as noted above, the superfluid vortex is not the
lowest energy configuration in the topological sector of configurations
with net winding number 1 at radial infinity. The lowest energy configuration 
consists of three (red, green, and blue)
semisuperfluid color flux tubes,
each with global winding $1/3$. A red semisuperfluid flux tube solution is
\bea \label{SemisuperfluidSoln}
\Phi_{\rm ssft}(r, \theta) &=& \bar\phi\,
  \left(\ba{ccc}f(r){\rm e}^{i \theta} & 0 & 0 \\
  0 & g(r) & 0 \\
  0 & 0 & g(r) \ea \right), \\[3ex]
A^{\rm ssft}_{\theta}(r) &=& - \dfrac{1}{gr} ( 1-h(r)) \left(\ba{ccc}-\frac{2}{3} & 0 & 0 \\
  0 & \frac{1}{3} & 0 \\
  0 & 0 & \frac{1}{3} \ea \right), \\[3ex]
 A^{\rm ssft}_{r} &=& 0 \ .
\eea
Green and blue flux tubes are obtained by permuting the diagonal elements.
The profile functions $f(r)$, $g(r)$ and  $h(r)$ obey 
\bea
&f^{''}& + \dfrac{f^{'}}{r} - \dfrac{(2h + 1)^{2}}{9 r^{2}} f - \dfrac{m_{\phi}^{2}}{6} f\left(f^2 + 2g^2-3 \right)\nonumber \\
&& - \dfrac{m_{\chi}^{2}}{3} f(f^2 -g^2) = 0 \ , \\
&g^{''}& + \dfrac{g^{'}}{r} - \dfrac{(h - 1)^{2}}{9 r^{2}} g - \dfrac{m_{\phi}^{2}}{6} g \left(f^2 + 2g^2-3 \right) \nonumber \\
&& + \dfrac{m_{\chi}^{2}}{6} g(f^2 -g^2) = 0 \ ,  \\
&h^{''}& - \dfrac{h^{'}}{r} 
 - \dfrac{m_{G}^{2}}{3} \left( g^{2}(h-1) + f^{2}(2h + 1) \right) = 0 \ ,
\eea
with  boundary conditions 
\bea
&&f \to 0,\ g'\to 0,\ h\to 1  \ \mathrm{as} \ r \to 0, \\
&&f \to 1,\   g \to 1,\ h \to 0 \ \mathrm{as} \ r \to \infty.
\eea

To understand why a configuration of three 
well-separated semisuperfluid flux tubes has lower energy than
a single superfluid vortex, compare the energy densities 
far from the core:
\beq
\ba{rcl}
\eps_{\rm sf} &=& 3 \bar\phi^2/r^2,  \\[1ex]
\eps_{\rm ssft} &=& \third \bar\phi^2/r^2.  
\ea
\label{eq:edens}
\eeq
%where $\eps_{\rm sf}$ and  $\eps_{\rm ssft}$ represent the energy densities of the superfluid vortex and the semisuperfluid flux tube respectively.
The energy density arises entirely from the scalar field gradient,
which for each
component is proportional to $n^2/r^2$, where $n$ is the net winding
of the field.
In the superfluid vortex there is net winding of 1 in each of the three
diagonal components, whereas in the semisuperfluid flux tube there is
net winding of $1/3$ in each component. So the energy density of a single semisuperfluid flux tube is $1/9$ of the energy density of a superfluid vortex.
This leads to a repulsive force between the flux tubes, since the further
apart they are, the more of space is filled with the energetically cheaper
semisuperfluid field configuration, as opposed 
to the energetically costlier superfluid vortex field configuration.
To estimate the leading term in the resultant potential
\beq
V(l) = E_{3\,\rm ssft} - E_{\rm sf} \ ,
\eeq
we note that the energy density of the three semisuperfluid 
vortices can be approximated as being the same as
a superfluid vortex at $r\gg l$, and being a superposition
of the three individual energy densities at $r\lesssim l$.
If we assume that the cores of the flux tubes have radius $\rho$, and
neglect the core contributions (which are independent of $l$), we find
\bea
V(l) &\approx&  2\pi \int_{\rho}^{l} \! r \mathrm{d}r \left(3 \eps_{\rm ssft} - \eps_{\rm sf}\right) \nonumber \\
% &=&   2 \pi \bar\phi^{2} \int_{\rho}^{l} \! r \mathrm{d}r \left( 3 \times \frac{1}{3 r^{2}} -  \frac{3}{r^{2}} \right) \nonumber \\
&=& - 4 \pi \bar\phi^{2} \ln(l/\rho) \ .
\eea 
For large separation $l$, there is a strong repulsive force decaying as
$1/l$. This justifies our neglect of contributions that would be subleading
in $l$, such as the core energies.

%============================================================
%  III: STABILITY ANALYSIS 
%============================================================
\section{Stability analysis of the superfluid vortex}
\label{sec:stability}

The Ginzburg-Landau energy functional for a two-dimensional
 static field configuration is 
\beq
E      = \int d^2 x \, \Hamil,
\eeq 
with the Hamiltonian density
\bea
\Hamil &=& \Tr \left[ \frac{1}{4} F_{ij} F^{ij} + D_{i} \Phi ^{\dagger} D^{i} \Phi   - m^2 \Phi ^{\dagger} \Phi  + \lambda_{2} (\Phi ^{\dagger} \Phi)^2 \right] \nonumber \\ 
&& + \lambda_{1}(\Tr[\Phi^{\dagger} \Phi])^2 + \dfrac{3m^{4}}{4 \lambda}, 
\eea
where $\lambda$ has been defined in equation (\ref{eq:lambda_def}).
We define the spatial behavior of the superfluid vortex \eqn{eq:superfluid_solution} by
\beq
\psi(r, \theta) \equiv  \bar\phi\beta(r) {\rm e}^{i \theta} \ .
\eeq
Up to an additive constant, the energy density of the vortex is
\beq
\Hamil(\Phi_{\mathrm{sf}}) = 3\left(|\partial_i \psi|^{2} - m^2|\psi|^2 + \lambda |\psi|^{4}\right) \ .
\eeq
To investigate the stability of the vortex, consider a perturbation
$\delta \Phi$ which only affects the quark condensate, leaving
the gauge field unperturbed,
\beq
\delta \Phi = \dfrac{f_0}{\sqrt{3}} \mathbf{1}_{3\times3} + f_{\alpha} T_{\alpha} \ .
\eeq
(We use the normalization $\Tr(T_{\alpha} T_{\beta}) = \delta_{\alpha \beta}$ for the $SU(3)$ generators.) 
To second order, the change in energy density due to the perturbation is given by,
\bea
&&\de\Hamil \equiv \Hamil(\Phi_{\mathrm{sf}} + \delta \Phi) - \Hamil(\Phi_{\mathrm{sf}}) = \delta \Hamil_{0} + \sum_{\alpha=1}^8 \delta \Hamil_{\alpha} \nonumber\\
&&= (|\partial_i f_0| ^2 + \lambda (\psi f_{0}^{*} + f_{0} \psi^{*})^{2} + |f_{0}|^{2}(-m^{2} + 2 \lambda |\psi|^{2})) + \nonumber  \\
&&\sum_{\alpha=1}^8 |\partial_i f_\alpha|^2 + \lambda_{2}(\psi f_{\alpha}^{*} + f_{\alpha} \psi^{*})^{2} + |f_{\alpha}|^{2}(-m^{2} + 2 \lambda |\psi|^{2}) \nonumber\\
&&+\mathcal{O}(f_{\{0,\alpha\}}^3) \ .
\eea
Focusing on perturbations in the $T_8$ color direction,
and decomposing $f_8=f_{8R} + i f_{8I}$, we can write
\beq
\ba{rcl}
{\cal \delta H}_8 &=& \left( f_{8R} \quad f_{8I}\right)
  \begin{pmatrix} \Om_{RR} & \Om_{RI} \\
  \Om_{RI} & \Om_{II} \end{pmatrix}
  \begin{pmatrix} f_{8R} \\ f_{8I} \end{pmatrix}, \\[2ex]
\Om_{RR} &=& -\nabla^2 + m^{2}\left(2 \dfrac{\lambda_2}{\lambda}  \beta^{2}\cos^{2}\theta +\beta^{2} -1 \right), \\
\Om_{II} &=& -\nabla^2 + m^{2}\left(2 \dfrac{\lambda_2}{\lambda}  \beta^{2}\sin^{2}\theta +\beta^{2} -1 \right), \\
\Om_{RI} &=& \dfrac{\lambda_2}{\lambda} m^{2} \beta^{2} \sin  2 \theta.
\ea
\label{eq:Omega}
\eeq
It can be shown that, for $\la_1>0$ (i.e.~$\la>\la_2$) the vortex 
is unstable to a perturbation of the form
\beq
\ba{rcl}
\delta \Phi^{(8)} &=&  \eps\, \hat n\!\cdot\!\nabla \psi(r,\th)\, T_8 \ , \\[1ex]
\ea
\label{eq:unstable-mode}
\eeq
which corresponds to a translation of the red and green
components of the vortex a small distance $\eps$
in the $\hat n$ direction, and translation
of the blue component a distance $2\eps$ in the opposite direction.

Using \eqn{eq:Omega} we obtain the energy of the
perturbation to order $\eps^2$,
\beq\label{eq:prediction}
\delta E_{8} =  \epsilon^{2} (\lambda_{2} -\lambda) \dfrac{\pi m^4 }{\lambda^{2}} \int_{0}^{\infty} \! r \mathrm{d}r \beta^{'2} \beta^{2} \ .
\eeq
This is the main result of this section. 
We see that if $\la>\la_2$ (i.e. $\la_1>0$) then the perturbation
\eqn{eq:unstable-mode} lowers the energy of the vortex.
At this point this is just a guess: there might be a lower-energy perturbation that
involves the gauge field or has a different spatial profile or
color structure. However, we will see in Sec. IV that the
numerically obtained unstable mode matches (23) very closely.

%============================================================
%  IV: NUMERICAL RESULTS 
%============================================================
\section{Numerical results}
\label{sec:results}

To analyze the instability of the superfluid vortex and map the
unstable/metastable boundary in the parameter space spanned by the three
couplings $g$, $\lambda$, and $\lambda_2$, we solve the classical field
equations on a two-dimensional lattice. For details see Appendix
\ref{sec:appendix_lattice_EOM}.
For all numerical calculations we chose the mass scale to be $m^2=0.25$ 
and use a lattice spacing $a=1$. 
This provides an adequate resolution for the superfluid vortex 
solution on the lattice for all parameter values that we studied,
since the size of the vortex depends only on $m^2$, not on any of the
couplings.

We use the ket $|\Phi\rangle$ as a convenient notation for the actual lattice configurations of matrices in 2-dimensional position space.
We define the following inner product and norm
\beq\label{eq:scalar_product}
\ba{rcl}
\langle A,B\rangle &\equiv & \int d^2x\mbox{Tr}\{B^\dagger A\} \ , \\[1ex]
||A|| &\equiv & \sqrt{\< A,A\>}\ .
\ea
\eeq
To characterize the degree to which field configurations resemble 
each other, we introduce the notion of an angle $\vartheta$ between 
two lattice configurations $A$ and $B$,
\beq\label{eq:angle}
\vartheta \equiv \arccos\frac{|\langle A,B\rangle|}{||A||\ ||B||} \ ,
% &=&\arccos\dfrac{|\int d^2x \mbox{Tr}\{B^\dagger A\}|}{\sqrt{\int d^2x \mbox{Tr}\{A^\dagger A\}}\sqrt{\int d^2x \mbox{Tr}\{B^\dagger B\}}}\nonumber.
\eeq
so when $\vartheta=0$ the configurations are the same up to an overall
multiplicative factor.

\subfiglabelskip=0pt
\begin{figure*}
\centering
\subfigure[][]{
 \label{fig:ednsty_a}
\includegraphics[width=0.45\hsize]{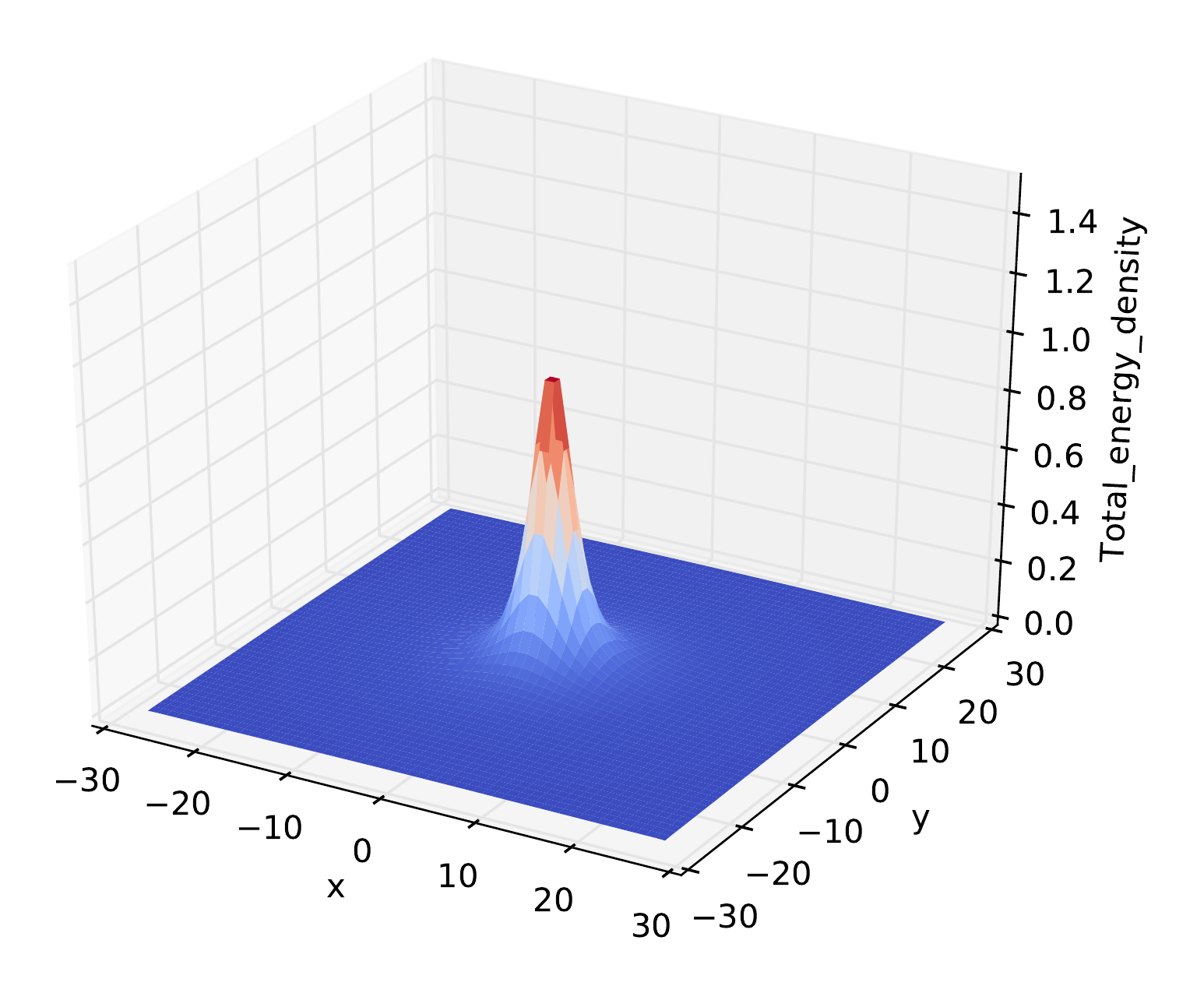}
}\hspace{8pt}
\subfigure[][]{
 \label{fig:ednsty_b}
\includegraphics[width=0.45\hsize]{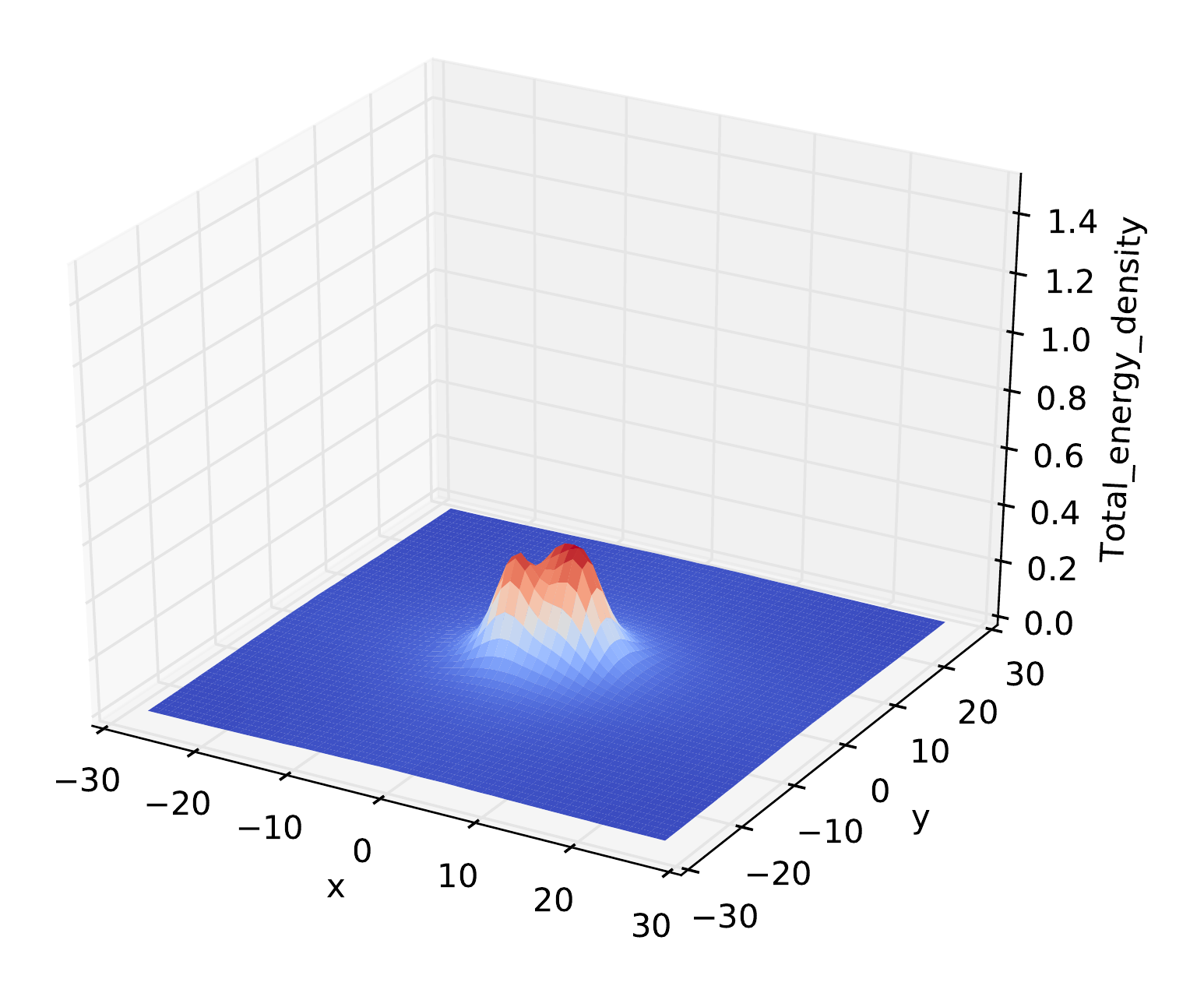}
}\\
\subfigure[][]{
 \label{fig:ednsty_c}
\includegraphics[width=0.45\hsize]{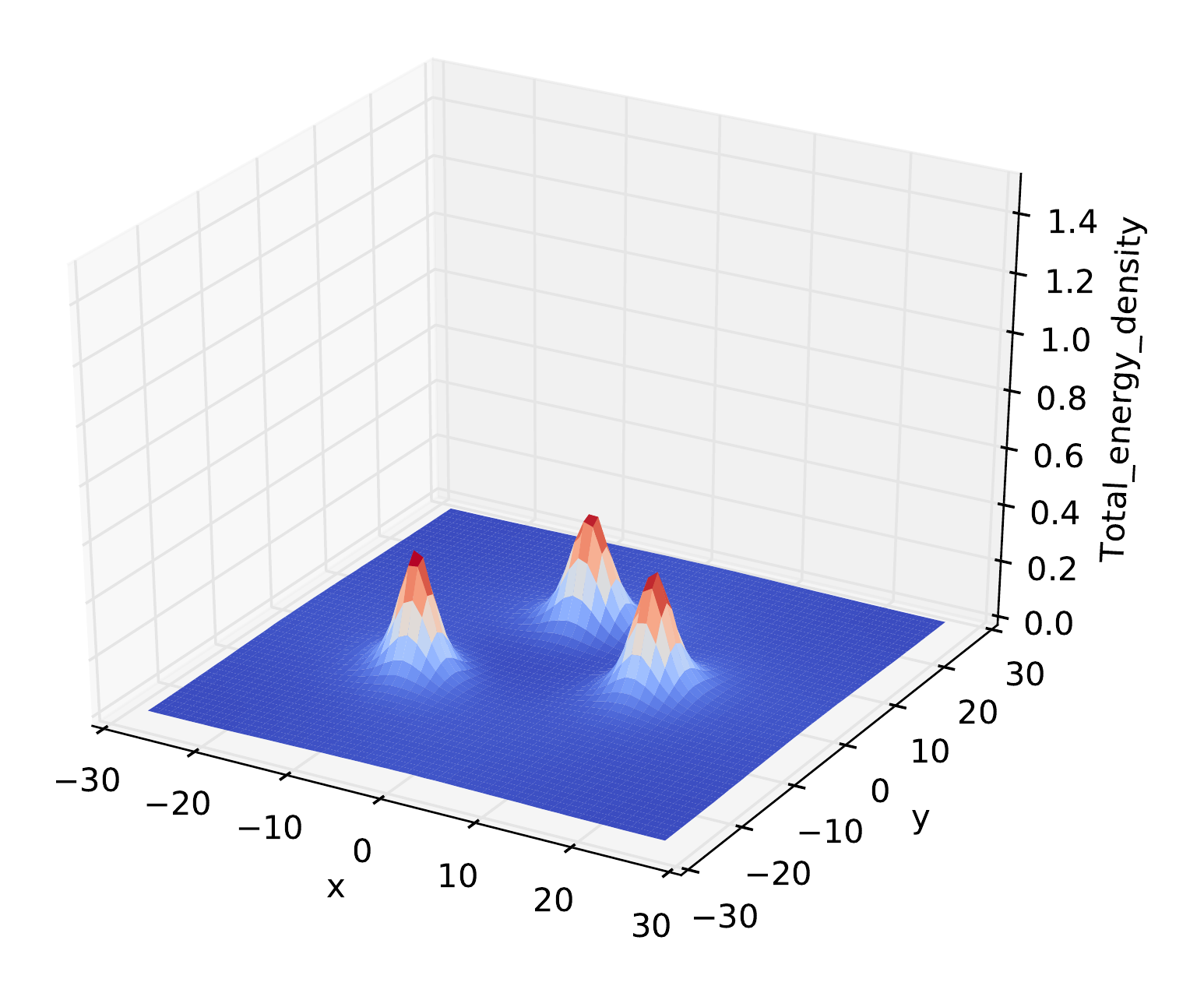}
}
\hspace{8pt}
\subfigure[][]{
 \label{fig:ednsty_d}
\includegraphics[width=0.45\hsize]{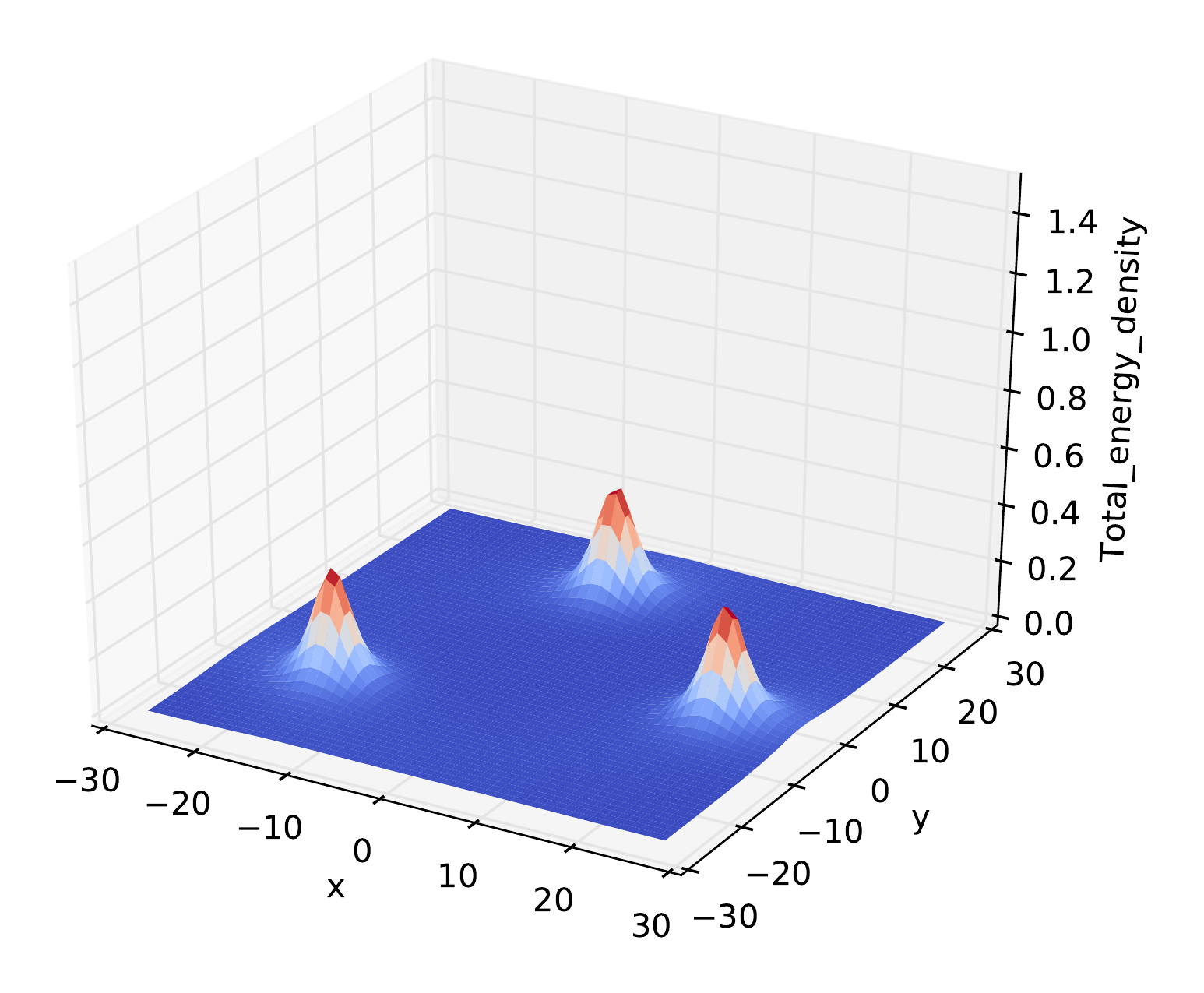}
}
\caption[]{Plots of the total energy density in position space for: \subref{fig:ednsty_a} a superfluid vortex before decay; \subref{fig:ednsty_b} fission of a superfluid vortex; \subref{fig:ednsty_c} formation of well separated semi-superfluid vortices; \subref{fig:ednsty_d} semi-superfluid vortices repelling each other.}
\label{fig:ednsty}
\end{figure*} 

\subsection{Numerical analysis of the unstable mode}

In parameter regions where the superfluid vortex is unstable, the
vortex solution has an unstable mode, and spontaneously decays to three
semisuperfluid flux tubes that repel each other.
Snapshots of this process are shown in Figure \ref{fig:ednsty}
where we plot the total energy density of the system along the
vertical axis.
The instability arises from
a direction in the high-dimensional configuration space 
of perturbations to the superfluid vortex
along which the potential possesses a negative curvature. This direction is the unstable mode $|\delta\Phi^{(u)}\>$,
and its time evolution is
\beq
\ba{rcl}
\partial_{tt}|\delta\Phi^{(u)}\rangle &=& \ga^2|\delta\Phi^{(u)}\rangle \ , \\[1ex]
|\delta\Phi^{(u)}\rangle_{t} &=& {\rm e}^{\ga t}|\delta\Phi^{(u)}\rangle_{t=0} \ .
\ea
\label{eq:unstable_diffeq}
\eeq 
In parameter regions where the vortex is metastable, there is no such
unstable direction, and all perturbations oscillate but remain small. 

To find out whether, for a given set of parameter values
$g$, $\lambda$, and $\lambda_2$, the superfluid
vortex is unstable or metastable, we proceed as follows.

We first generate a lattice field configuration $|\Phi_{\rm sf}\rangle$
that is the exact superfluid vortex solution of the
lattice field equations.
We do this by transferring the continuum vortex solution (with
radial profile obeying Eq.~(\ref{eq:radial_profile})) to the lattice
and then allowing it to relax to the lowest energy state by evolving
it using the Langevin equation with damping but no noise. 
We can do this even when the superfluid vortex is unstable because
the initial approximate vortex
configuration is proportional to the unit matrix in the color-flavor
space of complex $3\times 3$ matrices, and
the color-flavor symmetry of the Lagrangian guarantees that 
under time evolution it will remain proportional to the unit matrix,
whereas decay would require the generation of non-singlet color
components.

To probe the stability of the equilibrated vortex we
add a small perturbation $|\de\Phi^{(p)}\>$ to the superfluid vortex, and
evolve it forward in time.
The exact form of the perturbation is not important, as
long as it has some overlap with the unstable mode (if any).
If the vortex is unstable then even a tiny initial
perturbation with some component along the unstable direction
will grow exponentially as we evolve forward in
time. It quickly dominates the other components of the initial perturbation.
We tried three different perturbations for the $\Phi$-field (initial gauge links put to unity without perturbation):\\
\begin{enumerate}[(a)]
\item a random configuration\\ 
%\com{MGA: say something about how it was generated: normal distributions etc} \\
Here we used 8 complex random numbers (uniformly distributed between zero and $10^{-16}$) as entries for the $3\times3\ \Phi$-matrices at each lattice site.  
\item the analytically obtained unstable mode for $g=0$\\
This mode is constructed using (\ref{eq:unstable-mode}), which is controlled by two parameters, the direction $\hat n$ and magnitude of the displacement, $\epsilon$. Using the set $\{\textbf{1}_{3\times 3},T_\alpha\}$ as a complete basis, the radial profile serves as the position dependent coefficient of one or more of the basis elements $\{T_\alpha\}$, since $\textbf{1}_{3\times3}$ is a stable direction. We usually used $T_8$ only, but one can equally well choose any other basis element or combinations thereof.  
\item the numerically obtained unstable mode (see below).
\end{enumerate}

During an initial transient period $t_{\rm min}$
(which lasts longest for the
random initial perturbation) the unstable mode grows to become
the dominant component. After $t_{\rm min}$
the overlap of the growing mode with the original perturbation
grows exponentially in time, following \eqn{eq:unstable_diffeq},
\beq \label{eq:growth_random_mode}
A(t) \equiv \langle\delta\Phi^{(p)}|\Bigl(|\Phi\rangle_t-|\Phi_{\rm sf}\rangle\Bigr)
\propto {\rm e}^{\ga t},\quad
t_{\rm min} < t < t_{\rm max} \ .
\eeq
The exponential growth ends after time $t_{\rm max}$ when
the amplitude is so large that nonlinearities become non-negligible.
Using \eqn{eq:growth_random_mode} we can
measure the growth rate $\ga$ without knowing the actual form of the unstable mode. If such an exponential growth is observed, the superfluid vortex is unstable. 
The growth rate is determined 
by the negative curvature of the energy in the unstable direction,
and is independent of the initial ``seed'' perturbation.

For an unstable vortex we can numerically obtain the unstable mode
up to an overall normalization factor, by computing
\beq
|\delta\Phi^{(u)}\rangle\approx|\Phi\rangle_{t+\De t}-|\Phi\rangle_t \ .
\eeq
where $t_{\rm min} < t < t_{\rm max}$ and similarly for
$t+\De t$.
Once we know the unstable mode $|\delta\Phi^{(u)}\rangle$, we can go back and repeat the procedure described above 
using that mode as the initial seed perturbation,
in which case the initial transient time $t_{\rm min}$ is very short and
exponential growth starts immediately.

\begin{figure}
\centering
\includegraphics[width=\hsize]{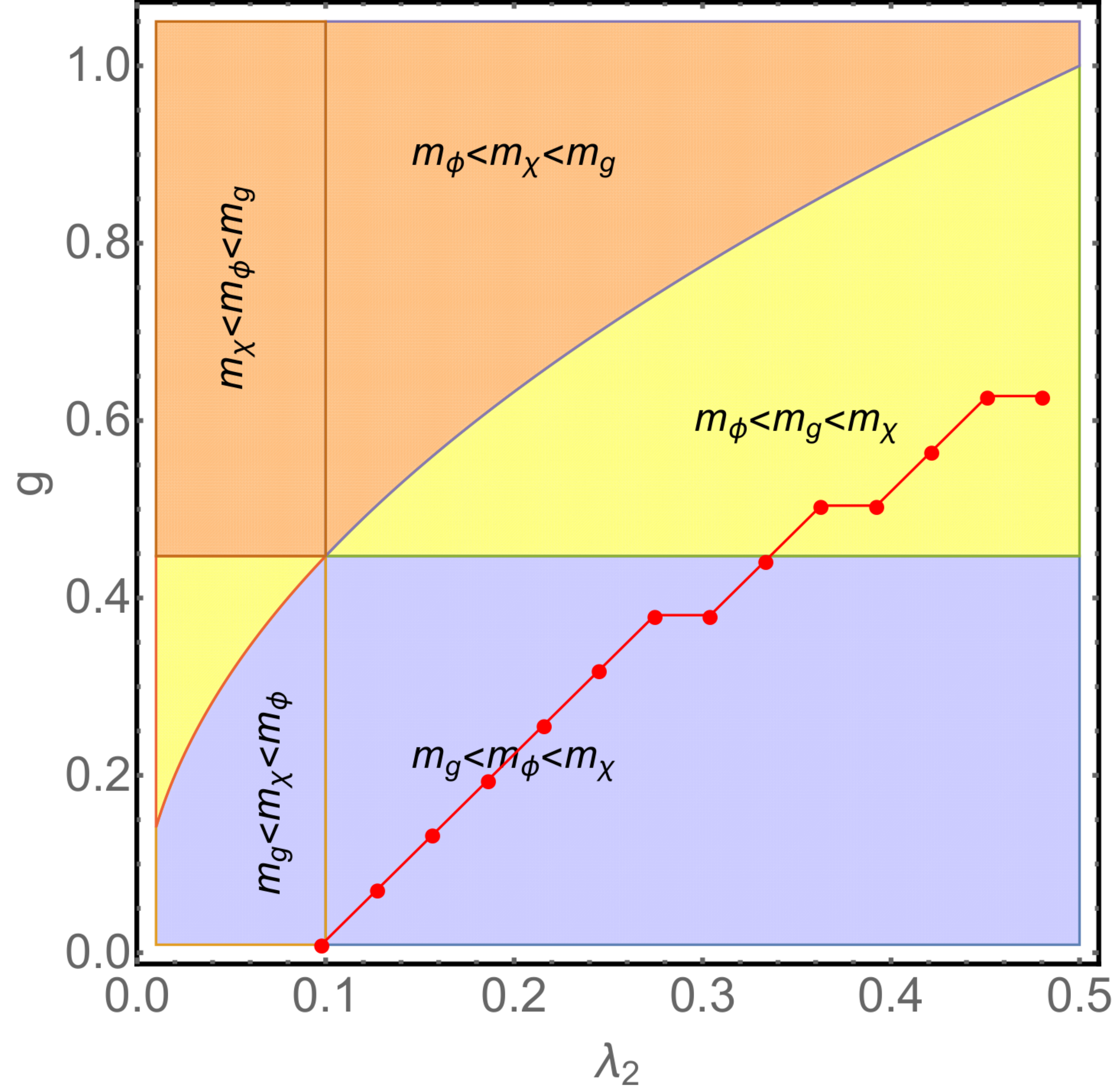}
\caption{Mass hierarchy in $g-\lambda_2$ plane for $\lambda=0.1$. The different shaded regions correspond to the different mass hierarchies of the scalar and gauge field masses. The intersection of the vertical line ($\lambda_{2} = \lambda $) with the horizontal axis is the point where the superfluid vortex changes from being unstable to being metastable at zero gauge coupling. The red points (connected by straight lines for better visual reference) denote the line along which the stable/unstable transition occurs. Points to the left of this line correspond to unstable configurations, points to the right to metastable superfluid solutions. The non-smooth nature of this line can be attributed to the finite resolution used while exploring the space of couplings.}
\label{fig:mass_ordering}
\end{figure}

During the epoch of exponential growth we use \eqn{eq:angle} to
compute the angle $\th$ between the growing mode and the initial perturbation.
When using the numerically constructed unstable mode as a seed we 
find that $\th$ is very small, typically in the range $10^{-2}$ to 2
degrees. When using an unstable mode that has been constructed analytically by Eq.~(\ref{eq:unstable-mode}), $\th$ is typically of the order of $10^{-3}$ degrees. 
Thus, the analytic mode discussed in Section (\ref{sec:stability}) seems to be identical to the unstable mode, at least as far as the $g=0$ case is concerned. Allowing for (larger) non-zero values of the gauge-coupling changes the picture, as indicated by Figure \ref{fig:mass_ordering}. The transition line between unstable and metastable configurations is not parallel to the $g$ axis, suggesting that the gauge field plays a more prominent role in the decay process as $g$ grows larger.
The angle in the case of the random mode is slightly larger as compared to the case where we used the analytic mode to perturb the system. This discrepancy is most likely a numerical artifact, as we are operating with very small numbers to isolate and extract the unstable mode in this very high-dimensional space.

\subsection{Parameter space scan}
\begin{figure*}
\bc
\includegraphics[width=0.9\hsize]{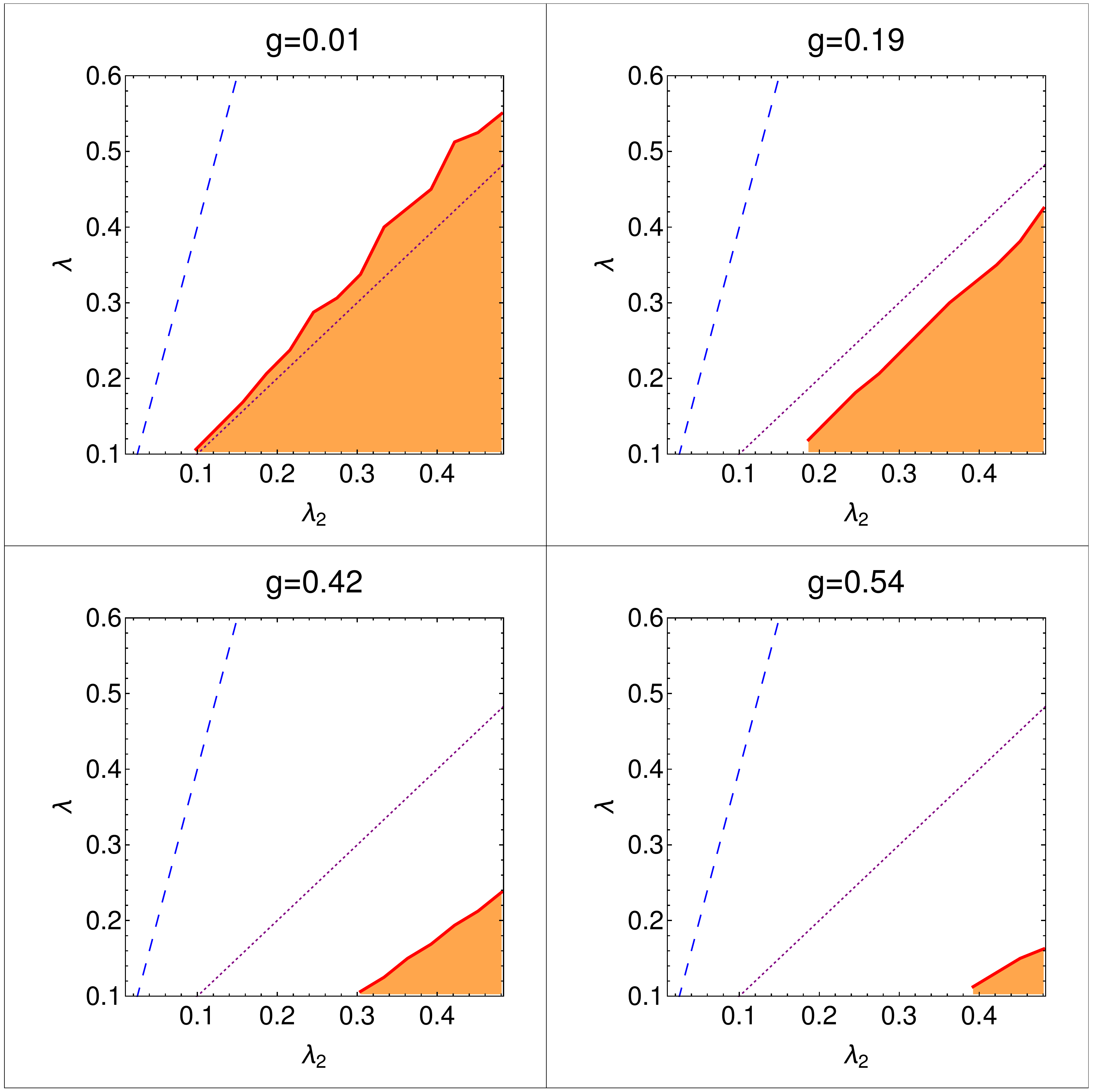}
\caption{Stability of the superfluid vortex for different values of $g$, $\lambda$ and $\lambda_2$. White areas indicate unstable sectors, while shaded areas correspond to regions where the superfluid vortex has been found to be stable. The dashed line corresponds to the line $\lambda=4\lambda_2$, along which the equations that relate the couplings to physical values of $\mu$ and $T_C$ are valid, see discussion in Section \ref{sec:mass_ratios}. According to the stability analysis discussed in Section \ref{sec:stability}, the transition region between stable and unstable solutions should correspond to the dotted line $\lambda=\lambda_2$ in the case of vanishing coupling $g$. This prediction seems to hold approximately in the case of small $g$.}   
\label{fig:parameter_space_contour_plot} 
\ec 
\end{figure*}

Using the procedure described above to determine whether the
vortices for a gives set of couplings $(g,\la,\la_2)$ are unstable
or metastable, we performed a scan of the parameter space
for $g\in[0.01,1]$, $\lambda\in[0.1,6]$, $\lambda_2\in[0.01,0.5]$.
The translation of these couplings to physical values is discussed in Section \ref{sec:mass_ratios} below. Our main findings are summarized in Figures \ref{fig:parameter_space_scan_plot} and \ref{fig:parameter_space_contour_plot}.

\begin{figure}
\bc
\includegraphics[width=\hsize]{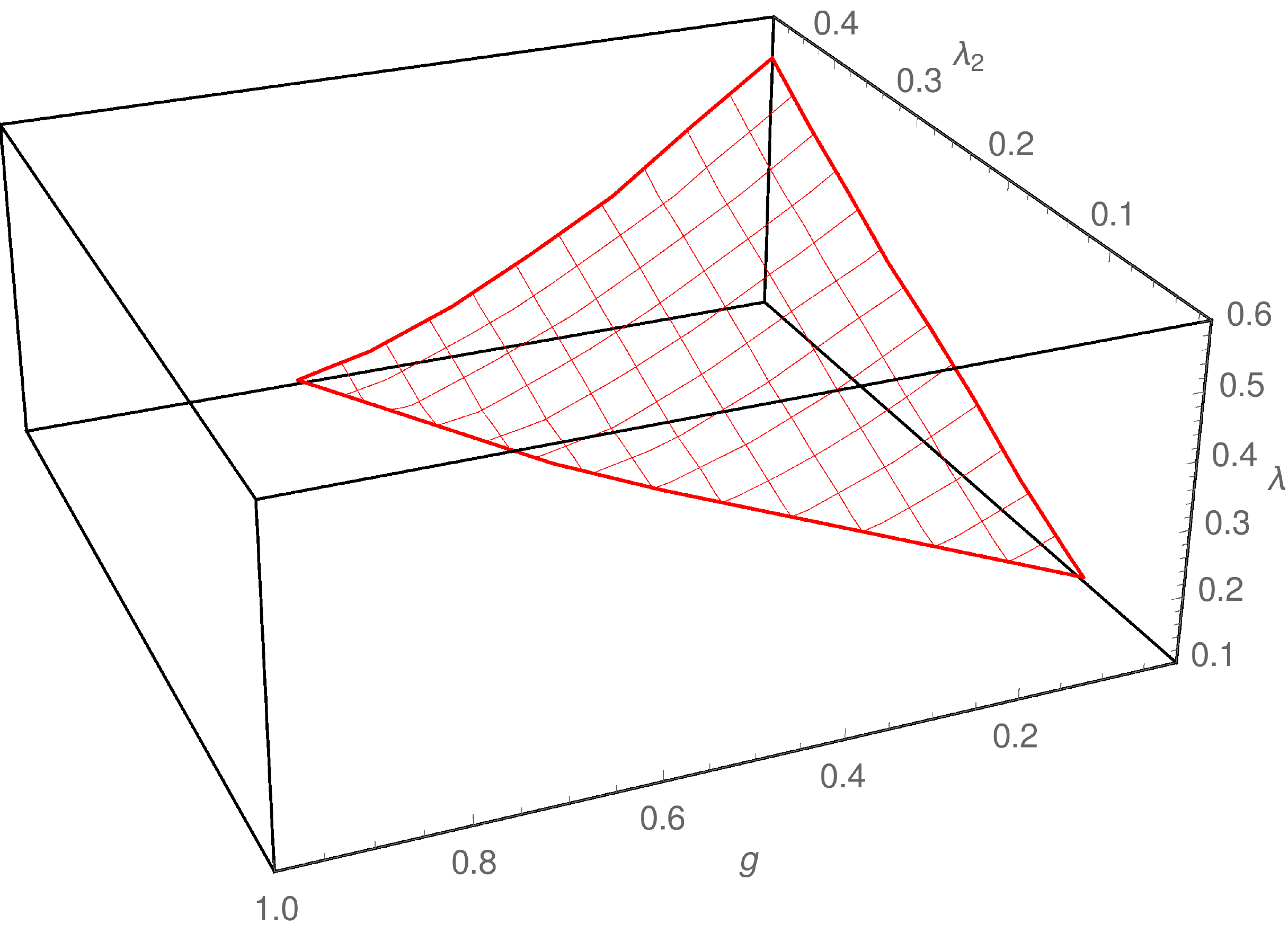}
\caption{The critical surface in the parameter space of the couplings $g$, $\lambda$ and $\lambda_2$. The surface separates the metastable from the unstable regime, where the latter corresponds to the larger volume shown in the plot. An exploration in $\lambda$-direction up to a value of $\lambda=6.0$ revealed nothing but unstable points.}       
\label{fig:parameter_space_scan_plot} 
\ec 
\end{figure}

The critical surface separating the metastable from the unstable regime is depicted in Figure~\ref{fig:parameter_space_scan_plot}.
We see that only a small subspace of the surveyed parameter space
yields metastable superfluid vortices. 
The plot suggests that at higher values of the gauge coupling ($g\gtrsim 1$),
which is the relevant region for QCD, the superfluid vortices are unstable,
spontaneously decaying in a short time.
Larger values of the coupling $\lambda$ also yield unstable vortices.

In Figure \ref{fig:parameter_space_contour_plot} we sliced the three-dimensional parameter space along the physically most relevant coupling among the three, the gauge coupling $g$. With increasing $g$, the area of metastable solutions decreases rapidly. The dotted line in the plot corresponds to $\lambda=\lambda_2$, which is the line along which the analytic analysis of Section \ref{sec:stability} predicts the stable/unstable transition. For small values of $g$, this seems to be approximately true. As the gauge coupling increases, however, the transition boundary deviates more and more from the predicted line, which is most likely due to the gauge field playing an increasingly important role in the decay process. 

\subsection{Relating stable and unstable regions to ratio of masses}
\label{sec:mass_ratios}

In Ref.~\cite{Eto:2009kg} it was suggested that the short range interaction
between semi-superfluid flux tubes, and hence the unstable/metastable boundary
for the superfluid vortex, is
dictated by the hierarchy of the masses $m_\phi$, $m_\chi$ and $m_g$ (Section \ref{sec:vortex_solutions}). In Figure \ref{fig:mass_ordering} we 
investigate this proposal by choosing
one slice of the parameter space along the plane $\lambda=0.1$
and plotting the different mass hierarchies along with the
unstable/metastable boundary.
The vertical line corresponds to the line along which $\lambda=\lambda_2$. We could not find a direct agreement between the unstable/metastable boundary and any of the boundaries derived from arranging the masses according to their mass hierarchy.

\subsection{Relation between the effective theory and QCD}

At arbitrary density it is not possible to calculate the
couplings in our effective theory in terms of the 
microscopic physics, namely QCD, because QCD is strongly coupled.
However, in the ultra-high density regime, where the coupling
becomes weak,
the parameters $\lambda_{1}$ and $\lambda_{2}$ have been calculated using the mean field approximation \cite{Iida:2000ha}, \cite{Giannakis:2001wz} in terms of baryon chemical potential $\mu$ and 
the transition temperature $T_{c}$ of the CFL condensate,
%(which can be calculated using the expressions from equations 2.10-2.13 from \cite{Eto:2013hoa}) 
\beq \label{LambdaExpressions}
\lambda_{1} = \lambda_{2} = \dfrac{\lambda}{4} = \dfrac{36}{7} \dfrac{\pi^4}{\zeta(3)} \left(\dfrac{T_{c}}{\mu} \right) ^{2}. 
\eeq 
Using this expression,
the range of values of $\la$ and $\la_2$ that we have explored in our
study correspond to values of $\mu$ ranging from 400 MeV to 500 MeV and values of $T_{c}$ ranging from 10 MeV to 15 MeV. As we can see, in the weak coupling mean field calculation $\lambda_{1} = \lambda_{2}$,
i.e. $\la=4\la_2$.
In our calculations the
superfluid vortex is unstable for $\lambda = 4 \lambda_{2}$.
We illustrate this in Figure \ref{fig:parameter_space_contour_plot}
where the dashed straight line is where $\la=4\la_2$, and even at very small
QCD coupling $g$ this line is in the unstable region.
Increasing the value of the coupling constant $g$ shrinks the region of meta-stability away from the plane  $\lambda = 4 \lambda_{2}$, and increasing the
value of $T_c$ just takes us to larger $\la$ and $\la_2$, so from
Figure \ref{fig:parameter_space_contour_plot} it seems likely that 
the vortices will remain unstable at large $g$ and $T_c$.

%============================================================
%  V: CONCLUSION
%============================================================
\section{Conclusion and discussion}
\label{sec:conclusion}

We have studied the stability properties of the superfluid vortices
in the CFL phase of dense quark matter.
Using a Ginzburg-Landau effective theory of the condensate field
discretized on a two-dimensional spatial lattice, we 
evolved the vortex configuration in time and looked for
an exponentially growing unstable mode.
We scanned the parameter space of the couplings, mapped the
regions where the vortices are unstable as opposed to
metastable, and in unstable regions we identified the unstable
mode.
We found that the region where superfluid vortices are metastable is rather small. Vortices are metastable when the gauge coupling $g$ is sufficiently small. In that case, the transition line separating the metastable from the unstable direction is almost as predicted in equation (\ref{eq:prediction}), that is, $\lambda\approx\lambda_2$, see upper left panel in Figure \ref{fig:parameter_space_contour_plot}. At larger values of $g$, the metastable region shrinks and vanishes around $g=0.6$. We could not find any sign of metastability above values of $\lambda\approx 0.6$.  
If we use mean-field weak-coupling calculations to relate the
Ginzburg-Landau couplings to QCD parameters such as $g$, $\mu$, and
$T$ then it seems likely that CFL vortices in neutron stars
would be unstable rather than
metastable, but these calculations are not valid in the density range
of interest for neutron stars. If better calculations of the
effective theory couplings become available, the physical region
in our plots could be more precisely identified.

It is very interesting that a superfluid vortex can be rendered unstable by solely perturbing the quark condensate,
in spite of the fact that
in CFL matter
gauge fields play an essential role in the later states of the decay process,
in which the vortex separates into three color flux tubes that repel each
other. 
In a theory without gauge fields there would be no such repulsive force,
and our preliminary calculations suggest that in that theory
the superfluid vortex decays into a molecule-like configuration of three separate vortices which remain
bound at a fixed spacial separation. 

This work opens up several avenues of future enquiry.
We only investigated the early stages of the process of vortex decay.
For regions of parameter space where the vortices are metastable it
would be interesting to measure their lifetime and evaluate the
height of the energy barrier.
Our Ginzburg-Landau theory did not include entrainment (current-current)
interactions, and it would be interesting to study how that affect our
results. The same methods that we used could be applied to
vortices in the color-spin-locked phase of quark matter, and to study the
stability of the proposed color-magnetic flux tubes in two-flavor
color superconducting quark matter.

%============================================================
%  ACKNOWLEDGMENTS
%============================================================
\begin{acknowledgments}
This material is based upon work supported by the U.S. Department of Energy, Office of Science, Office of Nuclear Physics under Award Number
% \#DE-FG02-91ER40628  % Wash U high energy theory
\#DE-FG02-05ER41375. AW acknowledges support by the Schr\"odinger Fellowship J 3800-N27 of the Austrian Science Fund (FWF). TV thanks Washington University for hospitality. TV acknowledges support by the U.S. Department of Energy, Office of Science, Office of High Energy Physics under
Award number \#DE-SC0013605.

\end{acknowledgments}

%============================================================
%  APPENDICES 
%============================================================
\appendix
\section{Equations of motion on the lattice}
\label{sec:appendix_lattice_EOM}

We solved the equations of motion on a two-dimensional spacial lattice whose lattice spacing we denote by $``a"$.
The matter field and its conjugate momentum
are represented by $3\times 3$ complex matrices $\Phi$ and $\Pi$
with one color 
and one flavor index which live on the lattice sites.
The gauge fields are $3\times 3$ unitary matrices $U$ with two color indices
living on the links, and their time derivative is encoded in the
electric field $E$ which is a Hermitian matrix. 
In what follows, $i$ and $j$ are discretized versions of spatial coordinates $x$ and $y$, and $t$ represents time. $U(t,i,j,+\hat{\mu})$ represents a gauge link at time $t$ that emanates from the site $(i,j)$ along the $\hat{\mu}$ direction. We perform our calculations in the temporal gauge, $A_{t} (t,i,j) = 0$.
One can construct a plaquette at time $t$, denoted by $U_\Box(t,i,j,\hat\mu)$, by starting at site $(i,j)$ and going in $\hat\mu$ direction. After each step, the orientation is changed by a 90-degree turn to the left, such that the plaquette closes after four steps. At any given time $t$, starting at site $(i,j)$, one can thus construct four different elementary plaquettes, depending on the direction of the initial step. The plaquette then corresponds to the product of the four link variables in the order of their appearance. For example, the plaquette $U_\Box(t,i,j,-\hat{y})$ becomes the matrix product $U^\dagger(t,i,j-1,+\hat{y})U(t,i,j-1,+\hat{x})U(t,i+1,j-1,+\hat{y})U^\dagger(t,i,j,+\hat{x})$.

The lattice energy functional is

\beq
E_{\mathrm{lattice}}  = \sum_{ij} (\Hamil_{\mathrm{magnetic}} + \Hamil_{\mathrm{electric}} + \Hamil_{\mathrm{potential}} + \Hamil_{\mathrm{kinetic}}),
\eeq
where
\bea
 \Hamil_{\mathrm{magnetic}} &=& \dfrac{6}{g^{2}a^{2}}\left(1 - \dfrac{1}{3} \Re\Tr \left[ U_{\Box}(t,i,j,+\hat{x}) \right]\right), \\ 
  \Hamil_{\mathrm{electric}} &=&  \dfrac{1}{2} \Tr \left[ E(t,i,j)^{2} \right], \\
 \Hamil_{\mathrm{potential}} &=&   \Tr \left[-m^{2} \Phi(t,i,j)^{\dagger} \Phi(t,i,j)\right]\\ 
&&+ \Tr\left[\lambda_{2} (\Phi(t,i,j) ^{\dagger} \Phi(t,i,j))^2  \right] \nonumber \\ 
&&+ \lambda_{1}(\Tr[\Phi (t,i,j) ^{\dagger} \Phi(t,i,j)])^2 + \dfrac{3m^{4}}{4 \lambda}, \nonumber \\
\Hamil_{\mathrm{kinetic}} &=&   \Tr \left[\Pi(t,i,j)^{\dagger} \Pi(t,i,j) \right].
\eea
The lattice equations of motion derived from the above energy functional are
\begin{widetext}
%\beq
%\nabla^{2}\Phi(t,i,j) \equiv  \dfrac{ \Phi(t,i+1,j) + \Phi(t,i-1,j) -2\Phi(t,i,j)}{a^2} + \dfrac{\Phi(t,i,j+1) + \Phi(t,i,j-1) -2\Phi(t,i,j)}{a^2}
%\eeq
\beq
\dot{\Pi}(t,i,j) = \nabla^{2}\Phi(t,i,j) + m^{2}\Phi(t,i,j) - 2 \lambda_{1} \Tr \left[ \Phi(t,i,j)^{\dagger} \Phi(t,i,j) \right] \Phi(t,i,j) 
                        - 2 \lambda_{2}\Phi(t,i,j) \Phi(t,i,j)^{\dagger}\Phi(t,i,j),
\eeq
where
%\beq
%\nabla^{2}\Phi(t,i,j) = \frac{1}{a^2}\sum_{\mu=x,y}\left( U(t,i,j,+\hat{\mu}) \Phi(t,i+1,j) + U^{\dagger}(t,i-1,j,+\hat{\mu}) \Phi(t,i-1,j) -2\Phi(t,i,j) \right),
%\eeq
\bea
\nabla^{2}\Phi(t,i,j) &=& \frac{1}{a^2}\left(U(t,i,j,+\hat{x}) \Phi(t,i+1,j) + U^{\dagger}(t,i-1,j,+\hat{x}) \Phi(t,i-1,j) -2\Phi(t,i,j) \right) \nonumber \\ 
&+& \frac{1}{a^2}\left( U(t,i,j,+\hat{y}) \Phi(t,i,j+1) + U^{\dagger}(t,i,j-1,+\hat{y}) \Phi(t,i,j-1) -2\Phi(t,i,j) \right),
\eea
\beq
 \dot{E}_{x}^{\alpha}(t,i,j) =
{\rm Im} \left(\Tr \left[T^\al \left(\dfrac{2}{g a} \left( U_{\Box} (t,i,j, +\hat{x}) - U_ {\Box} (t,i,j, -\hat{y}) \right) + \dfrac{2 g}{a} 
\left( 
U(t,i,j,+\hat{x}) \Phi(t,i+1,j) \Phi^{\dagger}(t,i,j) 
\right) \right) \right] \right) ,  
\eeq
\beq
 \dot{E}_{y}^{\alpha}(t,i,j) =
{\rm Im}\left( \Tr  \left[ T^\al \left( \dfrac{2}{g a} \left( U_{\Box} (t,i,j, +\hat{y}) - U_ {\Box} (t,i,j, +\hat{x} ) \right) + \dfrac{2 g}{a} 
\left( 
U(t,i,j,+\hat{y}) \Phi(t,i,j+1) \Phi^{\dagger}(t,i,j) 
\right) \right) \right]  \right) ,  
\eeq
\beq \label{eq:U}
U(t + \delta t,i,j,+\hat{\mu})= \exp\left[- \mathrm{i} g a\,  \delta t \,   E_{\mu}^{\alpha}(t+\delta t,i,j) T^{\alpha} \right]U(t,i,j,+\hat{\mu}),
\eeq
%\beq \label{eq:Uy}
%U(t + \delta t,i,j,+\hat{y})= \exp\left[- \mathrm{i} g a \, \delta t \,  E_{y}^{\alpha}(t+\delta t,i,j) T^{\alpha} \right]U(t,i,j,+\hat{y}),
%\eeq
where $T^{\alpha}$'s are the SU(3) generators. We choose the normalization $\Tr[T^{\alpha} T^{\beta}] = \delta^{\alpha \beta}$.
\beq \label{eq:Phi}
\Phi(t + \delta t,i,j) = \Phi(t,i,j) + \delta t \,  \Pi(t + \delta t,i,j) .
\eeq
\end{widetext}
In order to obtain the initial superfluid vortex configuration on the lattice, we use the Langevin evolution method. The continuum profile  of the superfluid vortex is first evolved using the Langevin approach, with temperature set to zero. We then take the final relaxed configuration and use it as the input configuration for our numerical studies on stability. In the Langevin implementation, equations (\ref{eq:U}) and (\ref{eq:Phi}) are modified as follows :
\begin{widetext}
\beq 
U(t + \delta t,i,j,+\hat{\mu})= \exp\left[- \mathrm{i} g a \, \delta t \, \left(E_{\mu}^{\alpha}(t,i,j) + \delta t \left(\dot{E}_{\mu}^{\alpha}(t,i,j) - \eta {E}_{\mu}^{\alpha}(t,i,j) +\zeta^{\alpha}(t,i,j)  \right) \right)T^{\alpha} \right]U(t,i,j,+\hat{\mu}),
\eeq
%\beq 
%U(t + \delta t,i,j,+\hat{y})= \exp\left[- \mathrm{i} g a \, \delta t \, \left(E_{y}^{\alpha}(t,i,j) + \delta t \left(\dot{E}_{y}^{\alpha}(t,i,j) - \eta {E}_{y}^{\alpha}(t,i,j) +\zeta^{\alpha}(t,i,j)  \right) \right)T^{\alpha} \right]U(t,i,j,+\hat{y}),
%\eeq
\beq
\Phi(t + \delta t,i,j) = \Phi(t,i,j) +  \delta t \left(\Pi(t,i,j) + \delta t\left(\dot{\Pi}(t,i,j) - \eta \Pi(t,i,j) + \zeta_{\mathrm{Re}}(t,i,j) + \mathrm{i}\,\zeta_{\mathrm{Im}}(t,i,j) \right) \right).
\eeq
\end{widetext}
where $\eta$ is the coefficient of viscosity. By the fluctuation-dissipation theorem the stochastic noise terms $\zeta_{\mathrm{Re}}$, $\zeta_{\mathrm{Im}}$, $\zeta^{\alpha}$ are independently drawn from the same Gaussian probability distribution
\beq
\zeta = \sqrt{\frac{2 \eta \Theta}{a ^{2} \delta t}} \,\xi(0,1)
\eeq 
where $\xi(0,1)$ is a Gaussian random number of zero mean and unit variance. $\Theta$ is the temperature of the thermal bath that is coupled to the system.

We tested two types of boundary conditions (BC): fixed and Neumann.
Fixed BC consisted of locking the matter and gauge fields
at the edge of the lattice to the values they would take
when there is a superfluid vortex
at the center of the lattice. With fixed BC the boundary affects
the later stages of 
the decay of a superfluid vortex because the edges repel
the semisuperfluid flux tubes. For Neumann BC we fixed the matter and gauge fields at the boundary to be
the same as their neighbors one lattice spacing in. This sets the gradient
of the field configuration to zero at the edge. With Neumann BC the
later stages of the decay behave correctly, with the three 
semisuperfluid flux tubes leaving the lattice and
disappearing across the boundary. However, with Neumann BC the
early states of the decay were affected by a very slight attraction
of the vortex to the boundary, so if the vortex lives for long enough
it starts moving slowly towards the  boundary. We therefore used
fixed BC in our calculations.

In order to study finite size effects, we used different lattice sizes for our calculations. Calculations that led to the main results of this paper were performed on a $61^2$ lattice and were repeated on a lattice of size $101^2$, where we found no significant discrepancy in the results. We furthermore varied the criteria in the code that are responsible for detecting the exponential growth of the unstable mode. We found our results to be robust under these changes as well.

%============================================================
%  BIBLIOGRAPHY
%============================================================

\bibliography{StabilityOfSuperfluidVortex_v1}{}

\begin{thebibliography}{10}%
\makeatletter
\providecommand \@ifxundefined [1]{%
 \@ifx{#1\undefined}
}%
\providecommand \@ifnum [1]{%
 \ifnum #1\expandafter \@firstoftwo
 \else \expandafter \@secondoftwo
 \fi
}%
\providecommand \@ifx [1]{%
 \ifx #1\expandafter \@firstoftwo
 \else \expandafter \@secondoftwo
 \fi
}%
\providecommand \natexlab [1]{#1}%
\providecommand \enquote  [1]{``#1''}%
\providecommand \bibnamefont  [1]{#1}%
\providecommand \bibfnamefont [1]{#1}%
\providecommand \citenamefont [1]{#1}%
\providecommand \href@noop [0]{\@secondoftwo}%
\providecommand \href [0]{\begingroup \@sanitize@url \@href}%
\providecommand \@href[1]{\@@startlink{#1}\@@href}%
\providecommand \@@href[1]{\endgroup#1\@@endlink}%
\providecommand \@sanitize@url [0]{\catcode `\\12\catcode `\$12\catcode
  `\&12\catcode `\#12\catcode `\^12\catcode `\_12\catcode `\%12\relax}%
\providecommand \@@startlink[1]{}%
\providecommand \@@endlink[0]{}%
\providecommand \url  [0]{\begingroup\@sanitize@url \@url }%
\providecommand \@url [1]{\endgroup\@href {#1}{\urlprefix }}%
\providecommand \urlprefix  [0]{URL }%
\providecommand \Eprint [0]{\href }%
\providecommand \doibase [0]{http://dx.doi.org/}%
\providecommand \selectlanguage [0]{\@gobble}%
\providecommand \bibinfo  [0]{\@secondoftwo}%
\providecommand \bibfield  [0]{\@secondoftwo}%
\providecommand \translation [1]{[#1]}%
\providecommand \BibitemOpen [0]{}%
\providecommand \bibitemStop [0]{}%
\providecommand \bibitemNoStop [0]{.\EOS\space}%
\providecommand \EOS [0]{\spacefactor3000\relax}%
\providecommand \BibitemShut  [1]{\csname bibitem#1\endcsname}%
\let\auto@bib@innerbib\@empty
%</preamble>
\bibitem [{\citenamefont {Alford}\ \emph {et~al.}(1998)\citenamefont {Alford},
  \citenamefont {Rajagopal},\ and\ \citenamefont {Wilczek}}]{Alford:1997zt}%
  \BibitemOpen
  \bibfield  {author} {\bibinfo {author} {\bibfnamefont {M.~G.}\ \bibnamefont
  {Alford}}, \bibinfo {author} {\bibfnamefont {K.}~\bibnamefont {Rajagopal}}, \
  and\ \bibinfo {author} {\bibfnamefont {F.}~\bibnamefont {Wilczek}},\ }\href
  {\doibase 10.1016/S0370-2693(98)00051-3} {\bibfield  {journal} {\bibinfo
  {journal} {Phys. Lett.}\ }\textbf {\bibinfo {volume} {B422}},\ \bibinfo
  {pages} {247} (\bibinfo {year} {1998})},\ \Eprint
  {http://arxiv.org/abs/hep-ph/9711395} {arXiv:hep-ph/9711395 [hep-ph]}
  \BibitemShut {NoStop}%
%%CITATION = HEP-PH/9711395;%%
\bibitem [{\citenamefont {Balachandran}\ \emph {et~al.}(2006)\citenamefont
  {Balachandran}, \citenamefont {Digal},\ and\ \citenamefont
  {Matsuura}}]{Balachandran:2005ev}%
  \BibitemOpen
  \bibfield  {author} {\bibinfo {author} {\bibfnamefont {A.~P.}\ \bibnamefont
  {Balachandran}}, \bibinfo {author} {\bibfnamefont {S.}~\bibnamefont {Digal}},
  \ and\ \bibinfo {author} {\bibfnamefont {T.}~\bibnamefont {Matsuura}},\
  }\href {\doibase 10.1103/PhysRevD.73.074009} {\bibfield  {journal} {\bibinfo
  {journal} {Phys. Rev.}\ }\textbf {\bibinfo {volume} {D73}},\ \bibinfo {pages}
  {074009} (\bibinfo {year} {2006})},\ \Eprint
  {http://arxiv.org/abs/hep-ph/0509276} {arXiv:hep-ph/0509276 [hep-ph]}
  \BibitemShut {NoStop}%
%%CITATION = HEP-PH/0509276;%%
\bibitem [{\citenamefont {Nakano}\ \emph {et~al.}(2009)\citenamefont {Nakano},
  \citenamefont {Nitta},\ and\ \citenamefont {Matsuura}}]{Nakano:2007dq}%
  \BibitemOpen
  \bibfield  {author} {\bibinfo {author} {\bibfnamefont {E.}~\bibnamefont
  {Nakano}}, \bibinfo {author} {\bibfnamefont {M.}~\bibnamefont {Nitta}}, \
  and\ \bibinfo {author} {\bibfnamefont {T.}~\bibnamefont {Matsuura}},\ }\href
  {\doibase 10.1016/j.physletb.2008.11.049} {\bibfield  {journal} {\bibinfo
  {journal} {Phys. Lett.}\ }\textbf {\bibinfo {volume} {B672}},\ \bibinfo
  {pages} {61} (\bibinfo {year} {2009})},\ \Eprint
  {http://arxiv.org/abs/0708.4092} {arXiv:0708.4092 [hep-ph]} \BibitemShut
  {NoStop}%
%%CITATION = ARXIV:0708.4092;%%
\bibitem [{\citenamefont {Eto}\ \emph {et~al.}(2014)\citenamefont {Eto},
  \citenamefont {Hirono}, \citenamefont {Nitta},\ and\ \citenamefont
  {Yasui}}]{Eto:2013hoa}%
  \BibitemOpen
  \bibfield  {author} {\bibinfo {author} {\bibfnamefont {M.}~\bibnamefont
  {Eto}}, \bibinfo {author} {\bibfnamefont {Y.}~\bibnamefont {Hirono}},
  \bibinfo {author} {\bibfnamefont {M.}~\bibnamefont {Nitta}}, \ and\ \bibinfo
  {author} {\bibfnamefont {S.}~\bibnamefont {Yasui}},\ }\href {\doibase
  10.1093/ptep/ptt095} {\bibfield  {journal} {\bibinfo  {journal} {PTEP}\
  }\textbf {\bibinfo {volume} {2014}},\ \bibinfo {pages} {012D01} (\bibinfo
  {year} {2014})},\ \Eprint {http://arxiv.org/abs/1308.1535} {arXiv:1308.1535
  [hep-ph]} \BibitemShut {NoStop}%
%%CITATION = ARXIV:1308.1535;%%
\bibitem [{\citenamefont {Cipriani}\ and\ \citenamefont
  {Nitta}(2013)}]{Cipriani:2013wia}%
  \BibitemOpen
  \bibfield  {author} {\bibinfo {author} {\bibfnamefont {M.}~\bibnamefont
  {Cipriani}}\ and\ \bibinfo {author} {\bibfnamefont {M.}~\bibnamefont
  {Nitta}},\ }\href {\doibase 10.1103/PhysRevA.88.013634} {\bibfield  {journal}
  {\bibinfo  {journal} {Phys. Rev.}\ }\textbf {\bibinfo {volume} {A88}},\
  \bibinfo {pages} {013634} (\bibinfo {year} {2013})},\ \Eprint
  {http://arxiv.org/abs/1304.4375} {arXiv:1304.4375 [cond-mat.quant-gas]}
  \BibitemShut {NoStop}%
%%CITATION = ARXIV:1304.4375;%%
\bibitem [{\citenamefont {Gleiser}\ and\ \citenamefont
  {Thorarinson}(2009)}]{Gleiser:2008dt}%
  \BibitemOpen
  \bibfield  {author} {\bibinfo {author} {\bibfnamefont {M.}~\bibnamefont
  {Gleiser}}\ and\ \bibinfo {author} {\bibfnamefont {J.}~\bibnamefont
  {Thorarinson}},\ }\href {\doibase 10.1103/PhysRevD.79.025016} {\bibfield
  {journal} {\bibinfo  {journal} {Phys. Rev.}\ }\textbf {\bibinfo {volume}
  {D79}},\ \bibinfo {pages} {025016} (\bibinfo {year} {2009})},\ \Eprint
  {http://arxiv.org/abs/0808.0514} {arXiv:0808.0514 [hep-th]} \BibitemShut
  {NoStop}%
%%CITATION = ARXIV:0808.0514;%%
\bibitem [{\citenamefont {Heinz}\ \emph {et~al.}(1997)\citenamefont {Heinz},
  \citenamefont {Hu}, \citenamefont {Leupold}, \citenamefont {Matinyan},\ and\
  \citenamefont {Muller}}]{Heinz:1996wx}%
  \BibitemOpen
  \bibfield  {author} {\bibinfo {author} {\bibfnamefont {U.~W.}\ \bibnamefont
  {Heinz}}, \bibinfo {author} {\bibfnamefont {C.~R.}\ \bibnamefont {Hu}},
  \bibinfo {author} {\bibfnamefont {S.}~\bibnamefont {Leupold}}, \bibinfo
  {author} {\bibfnamefont {S.~G.}\ \bibnamefont {Matinyan}}, \ and\ \bibinfo
  {author} {\bibfnamefont {B.}~\bibnamefont {Muller}},\ }\href {\doibase
  10.1103/PhysRevD.55.2464} {\bibfield  {journal} {\bibinfo  {journal} {Phys.
  Rev.}\ }\textbf {\bibinfo {volume} {D55}},\ \bibinfo {pages} {2464} (\bibinfo
  {year} {1997})},\ \Eprint {http://arxiv.org/abs/hep-th/9608181}
  {arXiv:hep-th/9608181 [hep-th]} \BibitemShut {NoStop}%
%%CITATION = HEP-TH/9608181;%%
\bibitem [{\citenamefont {Eto}\ and\ \citenamefont {Nitta}(2009)}]{Eto:2009kg}%
  \BibitemOpen
  \bibfield  {author} {\bibinfo {author} {\bibfnamefont {M.}~\bibnamefont
  {Eto}}\ and\ \bibinfo {author} {\bibfnamefont {M.}~\bibnamefont {Nitta}},\
  }\href {\doibase 10.1103/PhysRevD.80.125007} {\bibfield  {journal} {\bibinfo
  {journal} {Phys. Rev.}\ }\textbf {\bibinfo {volume} {D80}},\ \bibinfo {pages}
  {125007} (\bibinfo {year} {2009})},\ \Eprint {http://arxiv.org/abs/0907.1278}
  {arXiv:0907.1278 [hep-ph]} \BibitemShut {NoStop}%
%%CITATION = ARXIV:0907.1278;%%
\bibitem [{\citenamefont {Iida}\ and\ \citenamefont
  {Baym}(2001)}]{Iida:2000ha}%
  \BibitemOpen
  \bibfield  {author} {\bibinfo {author} {\bibfnamefont {K.}~\bibnamefont
  {Iida}}\ and\ \bibinfo {author} {\bibfnamefont {G.}~\bibnamefont {Baym}},\
  }\href {\doibase 10.1103/PhysRevD.63.074018, 10.1103/PhysRevD.66.059903}
  {\bibfield  {journal} {\bibinfo  {journal} {Phys. Rev.}\ }\textbf {\bibinfo
  {volume} {D63}},\ \bibinfo {pages} {074018} (\bibinfo {year} {2001})},\
  \bibinfo {note} {[Erratum: Phys. Rev.D66,059903(2002)]},\ \Eprint
  {http://arxiv.org/abs/hep-ph/0011229} {arXiv:hep-ph/0011229 [hep-ph]}
  \BibitemShut {NoStop}%
%%CITATION = HEP-PH/0011229;%%
\bibitem [{\citenamefont {Giannakis}\ and\ \citenamefont
  {Ren}(2002)}]{Giannakis:2001wz}%
  \BibitemOpen
  \bibfield  {author} {\bibinfo {author} {\bibfnamefont {I.}~\bibnamefont
  {Giannakis}}\ and\ \bibinfo {author} {\bibfnamefont {H.-c.}\ \bibnamefont
  {Ren}},\ }\href {\doibase 10.1103/PhysRevD.65.054017} {\bibfield  {journal}
  {\bibinfo  {journal} {Phys. Rev.}\ }\textbf {\bibinfo {volume} {D65}},\
  \bibinfo {pages} {054017} (\bibinfo {year} {2002})},\ \Eprint
  {http://arxiv.org/abs/hep-ph/0108256} {arXiv:hep-ph/0108256 [hep-ph]}
  \BibitemShut {NoStop}%
%%CITATION = HEP-PH/0108256;%%
\end{thebibliography}%
\bibliographystyle{apsrev4-1}
\end{document}